\documentclass[useAMS,usenatbib]{mnras}
\usepackage{graphicx}
\usepackage{epstopdf}
\DeclareGraphicsExtensions{.eps}

\newcommand{\Mwd}{\mbox{$M_{\mathrm{wd}}$}}
\newcommand{\Msun}{\mbox{$\mathrm{M}_{\odot}$}}
\newcommand{\Rsun}{\mbox{$\mathrm{R}_{\odot}$}}

\title[Doppler-imaging of a debris disc at a white dwarf]{Doppler-imaging of the planetary debris disc at the white dwarf SDSS\,J122859.93+104032.9}

\author[Christopher J. Manser et al.]{Christopher J. Manser\,$^1$\,{\Huge \footnotemark},
Boris T. G\"ansicke\,$^1$,
Thomas R. Marsh\,$^1$,
\newauthor Dimitri Veras\,$^1$,
Detlev Koester\,$^2$,
Elm\'e Breedt\,$^1$,
Anna F. Pala\,$^1$,
\newauthor Steven G. Parsons\,$^3$,
John Southworth\,$^4$\\
$^{1}$ Department of Physics, University of Warwick, Coventry CV4 7AL,
UK \\
$^{2}$ Institut f\"ur Theoretische Physik und Astrophysik, University of Kiel,
24098 Kiel, Germany\\
$^{3}$ Departamento de F\'isica y Astronom\'ia, Universidad de Valpara\'iso, Avenida Gran Bretana 1111, Valpara\'iso, Chile\\
$^{4}$ Astrophysics Group, Keele University, Staffordshire, ST5 5BG, UK
}

\begin{document}

\date{Accepted 20XX. Received 20XX; in original form 20XX}

\pagerange{\pageref{firstpage}--\pageref{lastpage}} \pubyear{20XX}

\maketitle

\label{firstpage}

\begin{abstract}
  
Debris discs which orbit white dwarfs are signatures of remnant planetary systems. We present twelve years of optical spectroscopy of the metal-polluted white dwarf SDSS\,J1228+1040, which shows a steady variation in the morphology of the 8600\,\AA\ Ca\,{\textsc{ii}} triplet line profiles from the gaseous component of its debris disc. We identify additional emission lines of O\,\textsc{i}, Mg\,\textsc{i}, Mg\,\textsc{ii}, Fe\,\textsc{ii} and Ca\,\textsc{ii} in the deep co-added spectra. These emission features (including Ca\,H\,\&\,K) exhibit a wide range in strength and morphology with respect to each other and to the Ca\,{\textsc{ii}} triplet, indicating different intensity distributions of these ionic species within the disc. Using Doppler tomography we show that the evolution of the Ca\,{\textsc{ii}} triplet profile can be interpreted as the precession of a fixed emission pattern with a period in the range 24--30\,years. The Ca\,{\textsc{ii}} line profiles vary on time-scales that are broadly consistent with general relativistic precession of the debris disc. 

\end{abstract}

\begin{keywords}
Stars: individual: SDSS\,J122859.93+104032.9 -- white dwarfs -- Circumstellar matter -- accretion, accretion discs -- line: profiles. 
\end{keywords}

\footnotetext{E-mail: C.Manser@warwick.ac.uk}

\section{Introduction}

The first discovery of a debris disc around a white dwarf was made by \cite{zuckerman+becklin87-1} via the detection of an infrared excess at G29--38, which was later interpreted as the emission from circumstellar dust by \cite{grahametal90-1}. This disc is thought to be the result of the tidal disruption of an asteroid or comet within a remnant planetary system that survived the main sequence evolution of the white dwarf progenitor \citep{jura03-1, debesetal12-1, verasetal14-1}.  Since the discovery of infrared emission from G29--38, debris discs have been detected at more than thirty additional white dwarfs \citep{kilicetal06-1, farihietal08-1, farihietal09-1, juraetal09-1, debesetal11-2, hoardetal13-1, bergforsetal14-1, rocchettoetal15-1}, which are in all (but possibly one, see \citealt{xuetal15-1}) cases accompanied by the detection of trace metals in the white dwarf photosphere \citep{koesteretal97-1, zuckermanetal07-1, jura+young14-1}.

Two decades after the discovery of dust at G29--38, \cite{gaensickeetal06-3} identified a gaseous disc around the white dwarf SDSS\,J122859.93+104032.9 (henceforth SDSS\,J1228+1040) via the detection of double peaked emission lines of Ca\,{\textsc{ii}} at 8498.02\,\AA, 8542.09\,\AA, 8662.14\,\AA, (henceforth the Ca\,{\textsc{ii}} triplet), which is indicative of Keplerian rotation in a flat disc \citep{horne+marsh86-1}. \textit{Spitzer} and \textit{HST} follow-up observations detected circumstellar dust and a plethora of metallic absorption lines, respectively; strengthening the connection of the gaseous material to the presence of a remnant planetary system around this white dwarf \citep{brinkworthetal09-1, gaensickeetal12-1}. The morphology of these emission lines provided firm dynamical confirmation that the debris disc resides within the tidal disruption radius of the white dwarf. 

Despite intense searches over the last decade, gaseous discs have been found around only seven white dwarfs \citep{gaensickeetal06-3, gaensickeetal07-1, gaensickeetal08-1, gaensicke11-1, farihietal12-1, melisetal12-1, wilsonetal14-1}. All seven also exhibit infrared excess from dusty discs, five of which were discovered after their gaseous counterpart \citep{dufouretal10-1, brinkworthetal12-1}. Only in HE\,1349--2305 and SDSS\,J095904.69--020047.6 (henceforth SDSS\,J0959--0200) were the Ca\,{\textsc{ii}} emission lines detected after the initial discovery of a dusty disc. Only a small fraction of the systems hosting a dust-dominated debris disc also have an observable gaseous component, implying that the presence of  detectable amounts of gas is not inherent to the presence of a debris disc.

The origin of these seven gaseous discs is still unclear. Several scenarios have been proposed, including runaway feedback between sublimation and the ensuing gas drag on the dust, leading to a rapid increase in the accretion rate onto the white dwarf \citep{rafikov11-2, metzgeretal12-1}, and secondary impacts of asteroids on an existing debris disc \citep{jura08-1}. Both cases involve dynamic interactions in the disc which might lead to detectable levels of variability.

Changes in the strength and morphology of the Ca\,{\textsc{ii}} triplet have been reported in a number of systems. SDSS\,J161717.04+162022.4 (\citealt{wilsonetal14-1}, henceforth SDSS\,J1617+1620) was identified to host a transient gaseous disc, which following its first detection monotonically decreased in strength over eight years until it dropped below detectability. In contrast SDSS\,J084539.17+225728.0 (\citealt{gaensickeetal08-1, wilsonetal15-1}, henceforth SDSS\,J0845+2257) shows changes in the morphology of the Ca\,{\textsc{ii}} triplet, but not in the strength of the lines. We also note that SDSS\,J0959--0200 has shown a large decrease in the infrared emission from the dusty component of the debris disc, however no time-resolved observations of the gaseous disc have been published \citep{farihietal12-1, xu+jura14-1}.

We have been spectroscopically monitoring the white dwarf SDSS\,J1228+1040 for twelve years and report in this paper pronounced changes in the morphology of the Ca\,{\textsc{ii}} triplet. We also show the first intensity map in velocity space of the Ca\,{\textsc{ii}} gas at SDSS\,J1228+1040 using Doppler tomography.

\begin{figure*}
\centerline{\includegraphics[width=2.1\columnwidth]{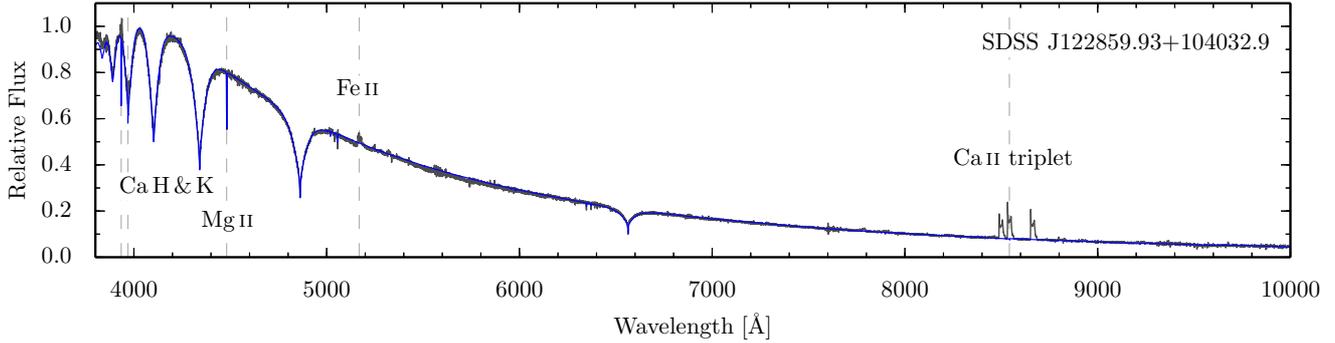}}
\caption{\label{f-1228_plot} X-shooter spectra (grey) of SDSSJ\,1228+1040 (obtained in January 2011) together with a model fit (blue) using the atmospheric parameters (Table\,\ref{t-params}). The strongest emission and absorption lines have been labelled.}
\end{figure*} 

\section{Observations}
\label{sec:observations}

\begin{table*}
\centering
\caption{Log of observations of SDSS\,J1228+1040. $^{a}$ Different exposure times for the individual UVES arms (blue / red lower and upper) and X-shooter arms (UVB / VIS). $^{b}$ Observations did not cover the Ca\,{\textsc{ii}} triplet. We did not use data collected by the NIR arm of X-shooter as the signal-to-noise ratio was too poor.  \label{t-dates}}
\begin{tabular}{rrrrr}
\hline
Date & Telescope/Instrument  & Wavelength Range [\AA] & Resolution [\AA] & Total Exposure Time [s]\,$^{a}$ \\
\hline
2003 March 27 & SDSS & 3800 -- 9200 & 2.7 & 2900\\
2006 June 30 - July 01 & WHT/ISIS & 7500 -- 9200 & 2.0 & 12000\\
2007 April 07 & VLT/UVES & 3045 -- 9463 & 0.19 & 11800 / 11360\\
2007 June 06 & VLT/UVES & 3758 -- 9463 & 0.18 & 5900 / 5680\\
2007 July 09 - 10 & VLT/UVES & 3757 -- 9463 & 0.18 & 5900 / 5680\\
2008 January 10 & VLT/UVES & 3759 -- 9464 & 0.18 & 5900 / 5680\\
2008 April 11\,$^{b}$ & VLT/UVES & 3284 -- 6649 & 0.13 & 11800 / 11360\\
2008 June 27 & VLT/UVES & 3758 -- 9463 & 0.18 & 5900 / 5680\\
2008 July 19 & VLT/UVES & 3758 -- 9463 & 0.18 & 5900 / 5680\\
2009 January 12\,$^{b}$ & VLT/UVES & 3284 -- 6650 & 0.14 & 5900 / 5680\\
2009 February 03 & VLT/UVES & 3758 -- 9464 & 0.18 & 5900 / 5680\\
2009 April 04 & VLT/UVES & 3282 -- 9463 & 0.18 & 17700 / 17040\\
2010 April 27 & WHT/ISIS & 8100 -- 8850 & 1.1 & 1800\\
2011 January 22 & VLT/X-Shooter & 2990 -- 10400 & 1.09 & 2840 / 1400\\
2011 May 29 & VLT/X-Shooter & 2990 -- 10400 & 1.14 & 2840 / 1400\\
2011 June 14 & VLT/X-Shooter & 2990 -- 10400 & 1.09 & 2840 / 1400\\
2012 March 26 & VLT/X-Shooter & 2990 -- 10400 & 1.11 & 2840 / 1400\\
2014 March 06 & VLT/X-Shooter & 2990 -- 10400 & 1.10 & 1971 / 3045\\
2014 June 02 & VLT/X-Shooter & 2990 -- 10400 & 1.09 & 1314 / 2030\\
2015 May 11 & VLT/X-Shooter & 2990 -- 10400 & 1.10 & 2380 / 2510\\ 
2015 May 18 & VLT/UVES & 3732 -- 9463  & 0.20 & 2980 / 2960\\
\hline
\end{tabular}
\end{table*}

\begin{table*}
\centering
\caption{Atmospheric parameters for SDSS\,J1228+1040 (from \citealt{koesteretal14-1}).   \label{t-params}}
\begin{tabular}{rrrr}
\hline
T$_{\mathrm{eff}}$\,[K] & $\log$\,g & M$_{\mathrm{WD}}$ [\Msun] & R$_{\mathrm{WD}}$ [\Rsun]\\
\hline
20713 (281) & 8.150 (0.089) & 0.705 (0.051) & 0.01169 (0.00078)\\
\hline
\end{tabular}
\end{table*}

\subsection{The data}

We obtained optical spectroscopy of SDSS\,J1228+1040 from 2003 to 2015 with several instruments: X-shooter \citep{vernetetal11-1} and the Ultraviolet and Visual Echelle Spectrograph (UVES, \citealt{dekkeretal00-1}) which are both on the ESO Very Large Telescope (VLT); the 2.5\,m Sloan Digital Sky Survey telescope (SDSS, data retrieved from DR7 and DR9, \citealt{gunnetal06-1, abazajianetal09-1, eisensteinetal11-1, ahnetal14-1, smeeetal13-1}); and the Intermediate dispersion Spectrograph and Imaging System (ISIS) on the William Herschel Telescope (WHT). These observations are summarised in Table\,\ref{t-dates}. The X-shooter and UVES data were reduced using the \texttt{\textsc{REFLEX}}\,\footnote{Documentation and software for \texttt{\textsc{REFLEX}} can be obtained from http://www.eso.org/sci/software/reflex/} reduction work flow using the standard settings and optimising the slit integration limits \citep{freudlingetal13-1}. The sky spectrum of each observation was used to determine the spectral resolution. We also report the parameters of SDSS\,J1228+1040 in Table\,\ref{t-params}. The first ISIS spectrum of SDSS\,J1228+1040 obtained on the WHT were reported in \cite{gaensickeetal06-3, gaensickeetal07-1}, the 2010 ISIS spectrum was obtained with a similar setup, and was reduced in the same fashion (see \citealt{farihietal12-1} and \citealt{wilsonetal14-1} for additional details).

\subsection{Telluric corrections}
The X-shooter observations (as with all ground-based spectroscopy) are affected by telluric lines, in particular around the Ca\,{\textsc{ii}} triplet. We removed this contamination using 152 empirical telluric templates from the X-shooter library (XSL), provided by \cite{chenetal14-1}. Each template includes the normalised telluric absorption features over the wavelength range 5300 -- 10200 \AA. The telluric absorption does not affect this entire range and hence we split the templates into seven smaller regions which we correct for independently. We first continuum normalise the observed spectrum and then we fit each of the seven regions to all 152 templates with two free parameters; a wavelength shift (for telescope flexure) and the intensity of the telluric features (which varies with the atmospheric conditions between different observations). The best-fitting template for each region is used to correct that part of spectrum. An example of a telluric corrected X-shooter spectrum for SDSS\,J1228+1040 is shown in Figure\,\ref{f-1228_plot}.

\subsection{Velocity corrections}

The observations of the Ca\,{\textsc{ii}} triplet were converted from wavelength to velocity space using the rest wavelengths, 8498.02\,\AA, 8542.09\,\AA, 8662.14\,\AA, which are corrected for the systemic velocity, i.e. the velocity along the line of sight. \cite{gaensickeetal12-1} calculated a velocity difference of $+57\,\pm\,1$\,km\,s$^{-1}$ between the photospheric absorption lines of the white dwarf and the interstellar ultraviolet absorption lines in SDSS\,J1228+1040. This value combines the systemic velocity as well as the gravitational red-shift of the photospheric absorption lines at the white dwarf surface \citep{koester87-1}.  Adopting a white dwarf mass and radius of $0.705\,\Msun$ and $0.01169\,\Rsun$, respectively \citep{koesteretal14-1}, we obtain a gravitational red-shift of $+38\,\pm\,4$\,km\,s$^{-1}$, resulting in a systemic velocity of $+19\,\pm\,4$\,km\,s$^{-1}$. This velocity is used to shift the Ca\,{\textsc{ii}} triplet observations to the rest frame of the system as well as for Doppler Tomography discussed in Section\,\ref{sec:doptom}. All the Ca\,{\textsc{ii}} triplet observations are shown in velocity space in Figure\,\ref{f-1228_Ca_vel} and we discuss the evolution of the Ca\,{\textsc{ii}} triplet in the next section.

\section{Evolution of the Ca\,{\textsc{ii}} Triplet structure}
\label{sec:evolution}

\subsection{Double peaked emission from a disc}

\cite{horne+marsh86-1} showed that a circular gaseous disc with a radially symmetric intensity distribution orbiting a central mass will produce symmetric double-peaked emission line profiles in velocity space (See Figure 1 of \citealt{horne+marsh86-1}), due to the range in velocities across the disc projected along the line of sight. Two properties of the physical structure of the disc, the inner and outer radii, can be easily inferred from these profiles as the Keplerian velocity increases with decreasing distance from the central object. The inner radius of the disc is determined from the maximum velocity, i.e. the point at which the emission drops to the continuum level; also known as the Full Width Zero Intensity (FWZI). We define the maximum red/blue-shifted velocities to represent the red/blue inner edges of the disc. The outer radius can be estimated from the peak separation in the double peaked profile. These observationally derived radii, $R_{\mathrm{obs}}$, relate to the true radii, $R$, via  $R_{\mathrm{obs}} = R\sin^{2}i$, where $i$ is the inclination of the disc. This method was used by \cite{gaensickeetal06-3} to derive the observed inner and outer radii of SDSS\,J1228+1040 to be $\simeq$\,0.64\,\Rsun\ and $\simeq$\,1.2\,\Rsun, respectively.

A departure from the symmetric double-peaked profile can reveal information about the physical structure of the disc, as demonstrated by \cite{steeghs+stehle99-1} for the occurence of spiral shocks in the disc \citep{steeghsetal97-1, steeghs99-1}. An eccentric disc would generate an asymmetric double-peaked profile if viewed along the semi-minor axis, as there is an asymmetry in the amount of material emitting red and blue-shifted light. We use this insight below to discuss the changes in the emission profiles seen in SDSS\,J1228+1040.

\begin{figure*}
\centerline{\includegraphics[width=2\columnwidth]{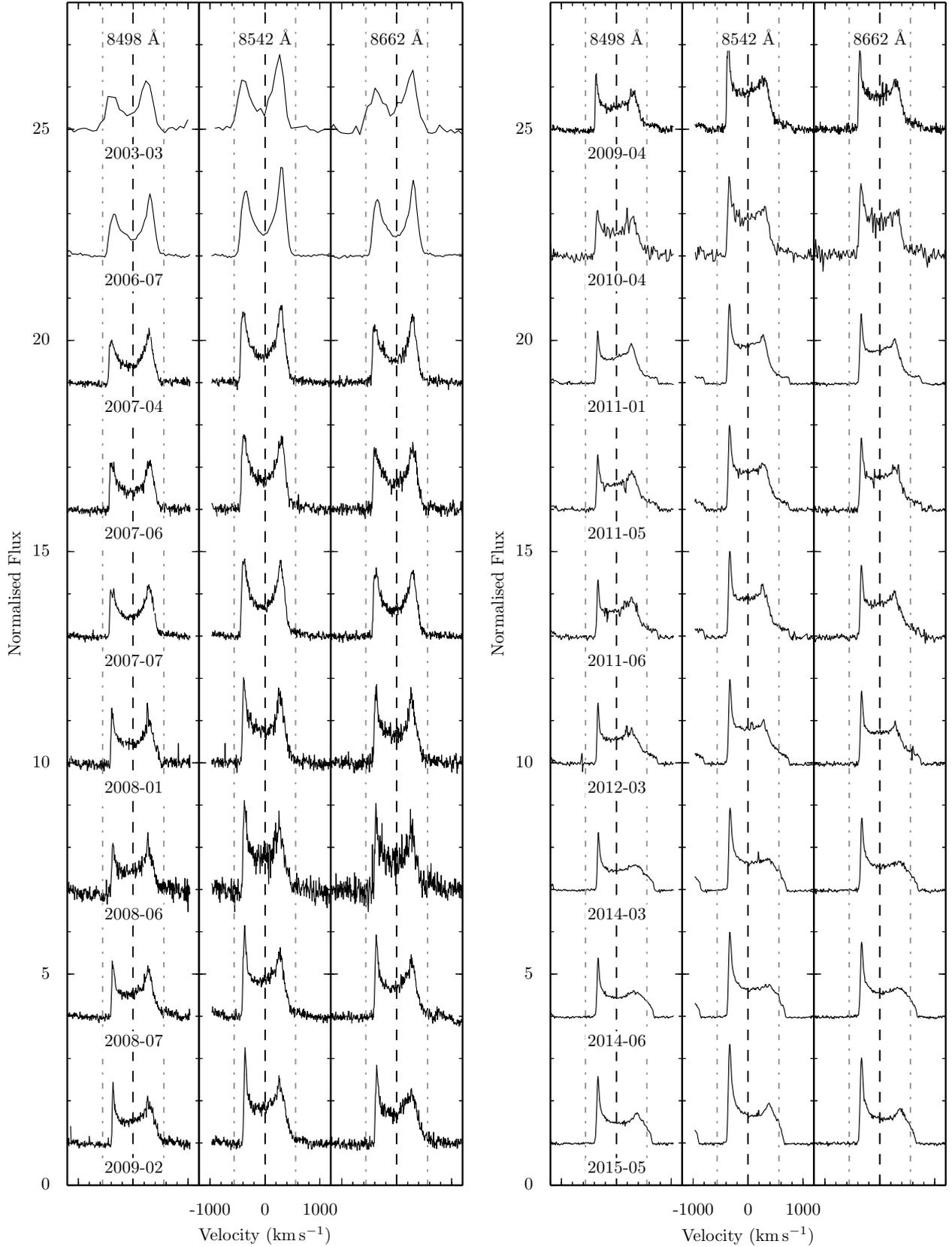}}
\caption{\label{f-1228_Ca_vel} Time series of the continuum-divided Ca\,{\textsc{ii}} triplet in SDSS\,J1228+1040 in velocity space. The series runs over twelve years with 18 epochs and depicts the change in the morphology of the line profiles. The dashed and dot-dashed lines indicate the zero velocity position, and the minimum and maximum velocities measured from the 2006 WHT data, respectively. The 2015 May profile is averaged from the X-shooter and UVES observations collected within several days of each other. The spectra are shifted in steps of three from the 2009 February (left) and 2015 May (right) observations.}
\end{figure*}

\subsection{Variation of the calcium triplet profiles}

The Ca\,{\textsc{ii}} triplet in SDSS\,J1228+1040 has undergone an astonishing morphological evolution in the twelve years of observations (Figure\,\ref{f-1228_Ca_vel}). The three individual line profiles vary in the same manner and we therefore use the term "profile" to decribe their evolution. The initial spectra from 2003 and 2006 show a double-peaked profile indicative of a circumstellar disc with a red-dominated asymmetry. These observations were interpreted as emission from an eccentric disc with a non isotropic intensity distribution \citep{gaensickeetal06-3}. The maximum velocities (highest blue or red-shift) in the disc measured from the 2006 observations were found to be $\pm\,560\,\pm\,10$\,km\,s$^{-1}$.

As the profiles evolve, there is a general smooth progression from a red-dominated asymmetry to a blue-dominated one. Throughout this evolution the morphology of the peaks also changes, with the blue-shifted peak becoming stronger and sharper. In contrast, the red-shifted peak becomes shallower and weaker and extends to higher velocities, which can be seen first in the April 2009 observations. This extension into the red is clearest in the January 2011 spectrum with a maximum extension of $+780\,\pm\,10$\,km\,s$^{-1}$. The clear, sharp cut-off of the red wing of the profile is a sudden change compared to the more gradual evolution in shape and implies that the red-shifted inner edge of the disc has moved inwards to smaller radii (i.e. higher velocities). From 2011--2015 the red wing decreases in extent, reducing the maximum velocity to $+670\,\pm\,10$\,km\,s$^{-1}$, accompanied by an increase in strength, showing a clear reappearance of the red peak. Overall, the profile shifts to longer wavelengths, with the blue edge moving from $-560\,\pm\,10$\,km\,s$^{-1}$ in July 2006 to $-390\,\pm\,10$\,km\,s$^{-1}$ in May 2015. The variation in the red and blue velocities imply an asymmetry in the inner disc edge, as a circular Keplerian orbit would produce identical red/blue velocities. In summary, the observed variations clearly show that the inner edge of the gaseous disc is not circular. 

\section{Additional Emission lines in SDSS\,J1228+1040}
\label{sec:add_emission}

\begin{figure*}
\centerline{\includegraphics[width=2\columnwidth]{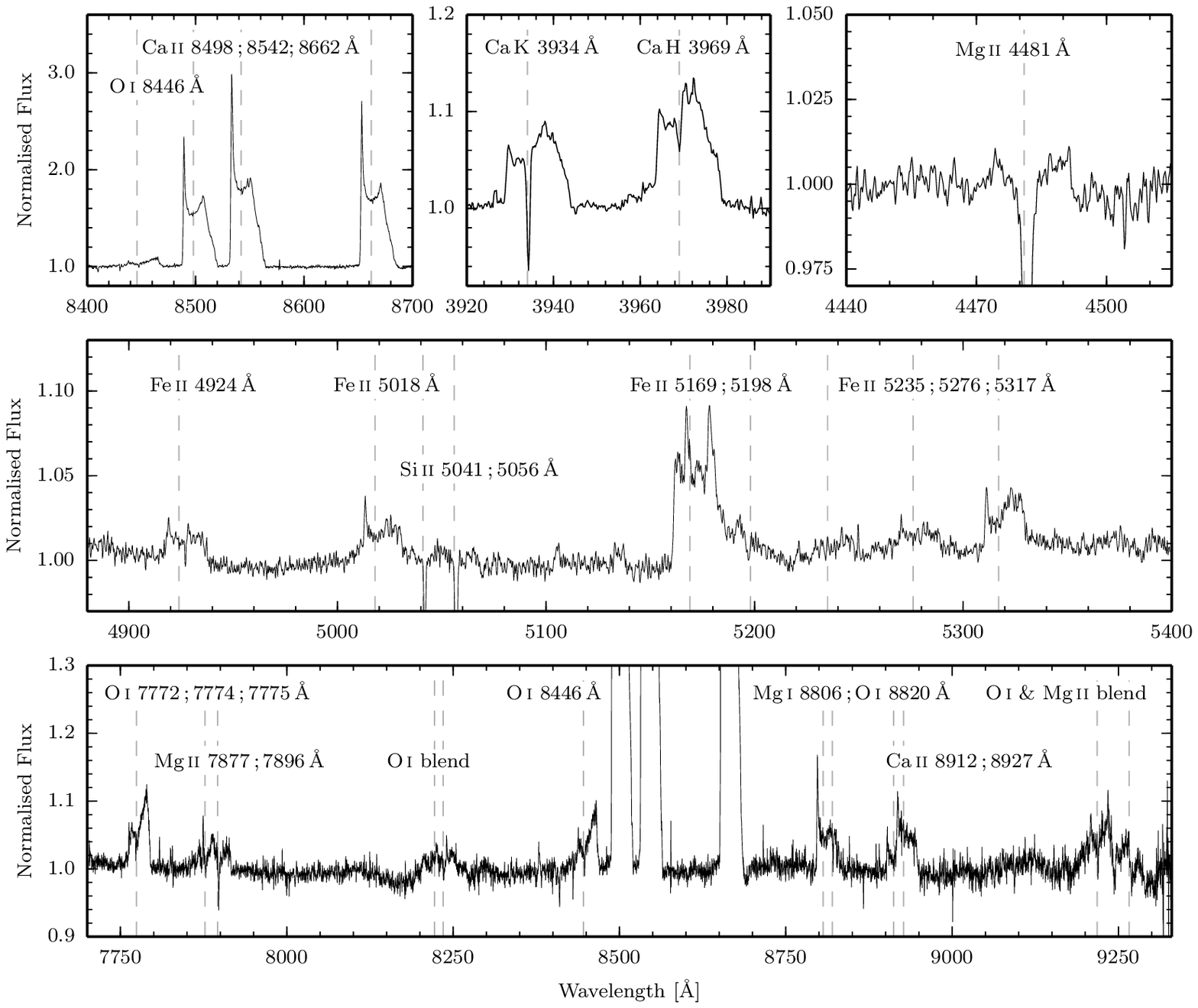}}
\caption{\label{other_emission} The averaged, continuum-divided X-shooter spectrum of SDSSJ\,1228+1040 reveals additional emission lines with a wide range of morphologies and strengths. The bottom panel contains the Ca\,{\textsc{ii}} triplet, which extends far off of the plot, illustrating the extremely large dynamical range of the data. The ions that contribute to each profile are labelled in each panel, with dashed lines indicating their rest wavelengths. The regions labelled as blends contain multiple lines (see Table\,\ref{t-lines}).}
\end{figure*}

In addition to the Ca\,\textsc{ii} triplet, \cite{gaensickeetal06-3} reported emission of Fe\,{\textsc{ii}} 5169\,/\,5197\,\AA, however these lines were too weak to resolve the shape of their profiles. By inspecting the averaged X-shooter UVB and VIS spectral arms, we clearly detected these Fe\,{\textsc{ii}} lines, as well as additional lines of Fe\,{\textsc{ii}}, Ca\,{\textsc{ii}}, Mg\,{\textsc{i}}, Mg\,{\textsc{ii}}, and O\,{\textsc{i}}. The rest wavelengths of these lines are listed in Table\,\ref{t-lines}, and their profiles are shown in Figure\,\ref{other_emission}. We cannot inspect the time-resolved evolution of these lines as they are very weak and correspondingly noisy in the individual spectra. 

\begin{table}
\centering
\caption{Additional emission lines detected in the average X-shooter spectrum (Figure\,\ref{other_emission}) and the \textit{HST} spectrum (Figure\,\ref{f-HST}) of SDSS\,J1228+1040). The rest wavelengths were obtained from the NIST Atomic Spectra Database. \label{t-lines}}
\begin{tabular}{ll}
\hline
Ion & Rest wavelengths [\AA] \\
\hline
Ca\,{\textsc{ii}} & 3933.66, 3968.47, 8498.02, 8542.09, 8662.14, 8912.07 \\
 & 8927.36\\
O\,{\textsc{i}} & 7771.94, 7774.17, 7775.39, 8221.82, 8227.65, 8230.02,\\ 
 & 8233.00, 8235.35, 8446.36, 8446.76, 8820.43, 9260.81,\\ 
 & 9260.84, 9260.94, 9262.58, 9262.67, 9262.77, 9265.94,\\  
 & 9266.01\\
Mg\,{\textsc{i}} & 8806.76\\
Mg\,{\textsc{ii}} & 2795.53, 2798.00, 2802.70, 4481.13, 4481.33, 7877.05, \\
 & 7896.37, 8234.64, 9218.25\\
Fe\,{\textsc{ii}} & 4923.92, 5018.44, 5169.03, 5197.57, 5234.62, 5276.00,\\
 & 5316.61, 5316.78 \\
\hline
\end{tabular}
\end{table}

It is evident that the four different species have distinct intensity distributions across the disc, which lead to the observed variety of line profile shapes. The sharp blue peak in the Ca\,{\textsc{ii}} triplet is observed only in Mg\,{\textsc{i}} 8806\,\AA\ and Ca\,{\textsc{ii}} 8912, 8927\,\AA. Rather surprisingly, the Ca H \& K lines are different in shape to the Ca\,{\textsc{ii}} triplet, with a broad profile, much weaker in strength, and no clear asymmetry in the blue edge. They do, however, possibly show an extension to the red similar to the Ca\,{\textsc{ii}} triplet, although the signal-to-noise ratio of the individual observations makes the presence of such an extension uncertain. The O\,{\textsc{i}} lines at $\simeq$\,7774 and 8446\,\AA\ show a clear red-dominated asymmetric profile, which has the opposite shape of the Ca\,{\textsc{ii}} triplet. The two emission profiles at $\sim$\,8245 (O\,{\textsc{i}}) and $\sim$\,9250\,\AA\ (O\,{\textsc{i}} and Mg\,{\textsc{ii}}) are blended, extending over a larger wavelength range with a blurred structure.

A number of emission profiles, such as Mg\,{\textsc{ii}} 7896\,\AA, show sharp absorption features. These features are the photospheric absorption lines, which is also the case at 5041 and 5056\,\AA, where a Si\,{\textsc{ii}} absorption doublet punctures the continuum.

\begin{figure*}
\centerline{\includegraphics[width=2\columnwidth]{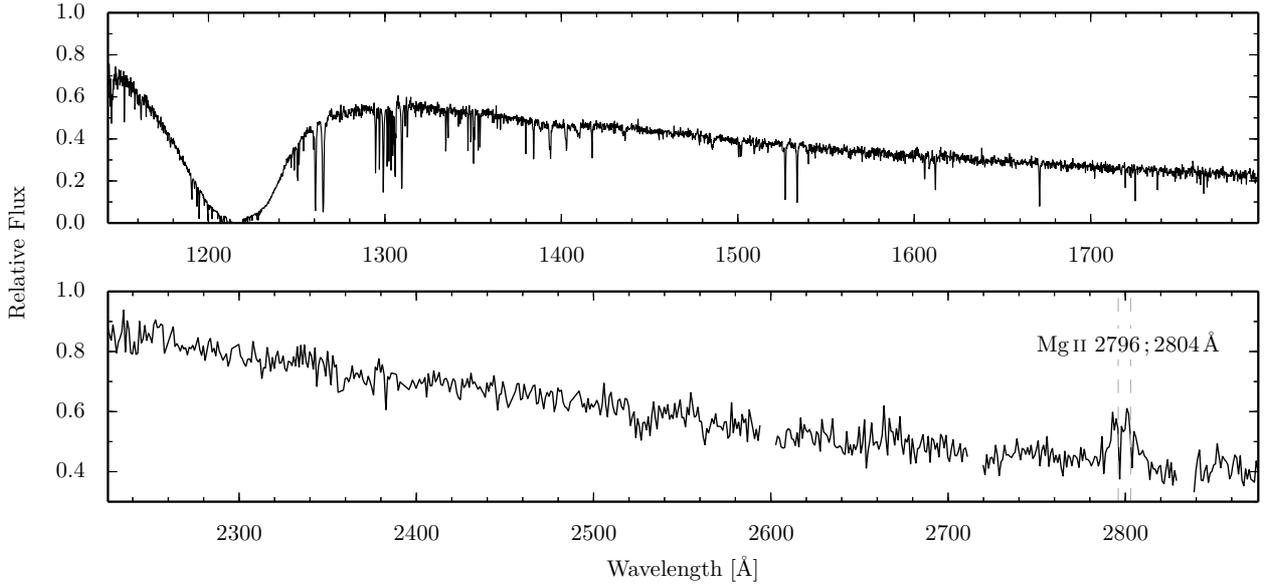}}
\caption{\label{f-HST} \textit{HST} far (top) and near (bottom) ultraviolet observations of SDSS\,J1228+1040. The Mg\,{\textsc{ii}} emission line is labelled.}
\end{figure*}

We also show for reference the \textit{HST} spectrum (Figure\,\ref{f-HST}) obtained by \cite{gaensickeetal12-1} in the far and near ultraviolet spectral range. There are far more photospheric absorption lines in the ultraviolet compared to the optical range, but only one emission line; Mg\,{\textsc{ii}} 2800\,\AA, which was first noted by \cite{hartmannetal11-1}. 

\begin{table}
\centering
\caption{Equivalent widths (EW) of the emission features shown in Figures\,\ref{other_emission}\,\&\,\ref{f-HST}. The Ca\,{\textsc{ii}} triplet are set in bold. \label{t-ew}}
\begin{tabular}{l c r}
\hline
Identifier & Central wavelength [\AA] & EW (error) [\AA] \\
\hline
Mg\,\textsc{ii} & 2800 & -7.9 (0.4) \\
Ca\,\textsc{ii} & 3934 & -0.893 (0.007) \\
Ca\,\textsc{ii} & 3969 & -1.485 (0.009) \\
Mg\,\textsc{ii} & 4481 & -0.061 (0.002) \\
Fe\,\textsc{ii} & 4923 & -0.243 (0.007) \\
Fe\,\textsc{ii} & 5018 & -0.377 (0.008) \\
Fe\,\textsc{ii} & 5185 & -1.60 (0.01) \\
Fe\,\textsc{ii} & 5235 & -0.220 (0.009) \\
Fe\,\textsc{ii} & 5276 & -0.478 (0.009) \\ 
Fe\,\textsc{ii} & 5317 & -0.646 (0.007) \\
O\,\textsc{i} & 7774 & -2.20 (0.02) \\
Mg\,\textsc{ii} & 7877 & -0.78 (0.02) \\
O\,\textsc{i} & 8230 & -0.71 (0.03) \\
O\,\textsc{i} & 8446 & -1.50 (0.02) \\
Ca\,\textsc{ii} & \textbf{8498} & \textbf{-16.73 (0.03)} \\
Ca\,\textsc{ii} & \textbf{8542} & \textbf{-23.45 (0.03)} \\
Ca\,\textsc{ii} & \textbf{8662} & \textbf{-20.76 (0.03)} \\
Mg\,\textsc{i}, O\,\textsc{i} & 8810 & -1.73 (0.03) \\
Ca\,\textsc{ii} & 8920 & -1.90 (0.03) \\
O\,\textsc{i}, Mg\,\textsc{ii} & 9234 & -2.64 (0.05) \\
\hline
\end{tabular}
\end{table}

As done above for the Ca\,{\textsc{ii}} triplet, we measured the maximum red/blue velocities of the emission profiles in the optical (where possible) to determine the radial extent of the different species. The additional emission features are orders of magnitude weaker than the  strength of the Ca\,{\textsc{ii}} lines (see Table\,\ref{t-ew}), and it is therefore difficult to accurately determine the maximum red and blue-shifted velocities. However, the profiles have a similar FWZI in velocity space and as such we report the average velocities of the blue-most and red-most edge of the profiles as $\simeq$\,400\,km\,s$^{-1}$ and $\simeq$\,800\,km\,s$^{-1}$, respectively, which are consistent with those obtained for the Ca\,{\textsc{ii}} triplet in all of the X-shooter observations so far. The similarity in the maximum and minimum velocities suggests that the emitting ions share a similar location of their inner disc edges. This is interesting to note as the line profiles differ in shape, implying different intensity distributions across the disc. Estimating an outer edge of the disc is much more difficult, as most of the additional emission profiles only have one clearly visible peak, and therefore a peak separation cannot be measured.

We compare the enlarged set of emission lines (Table\,\ref{t-lines}) to the predictions from Non-Local Thermodynamic Equilibrium (NLTE) models of gaseous discs around white dwarfs \citep{hartmannetal11-1}. Each of the Ca and Mg lines detected in our X-shooter spectra are also present in the model of \cite{hartmannetal11-1}, although they also predict other lines from these elements which we have not observed. These lines may lie below the detection threshold of our data, e.g. the predicted Mg\,{\textsc{i}} 5173\,\AA\ line is probably dominated by the emission of Fe\,{\textsc{ii}} 5169\,\AA. We cannot compare our detections of Fe\,{\textsc{ii}} or O\,{\textsc{i}} as these are not included in the calculations of \cite{hartmannetal11-1}.

\cite{hartmannetal11-1} predict the Ca\,{\textsc{ii} triplet to be flanked by emission of C\,{\textsc{ii}} 8700\,\AA. By comparison with the 2006 WHT spectra, the authors constrained the C abundance in the disc to less than 0.46\,per\,cent by mass. \cite{gaensickeetal12-1} measured the C abundance of the debris accreted by SDSS\,J1228+1040 from \textit{HST} UV spectra and found a mass fraction of 0.02\,per\,cent, consistent with the non-detection of C emission in our deeper X-shooter spectra.

\section{Doppler Tomography}
\label{sec:doptom}

While it is thought that the emission lines originate in a gaseous disc with an assumed Keplerian velocity field, the intensity distribution across the disc is not known. The smooth morphology variations of the Ca\,{\textsc{ii}} triplet imply a gradual change in that distribution as viewed from Earth over the twelve years of observations, which is astonishing when compared to the orbital period of the material in the disc, which is of the order of a few hours.

One way of inducing the observed line profile variation could be the rotation (precession) of a fixed non-axisymmetric intensity structure. This has been proposed to explain some of the features observed in the emission line profiles seen at Be stars. These stars are rapidly rotating and host circumstellar material in disc structures that frequently show long term variability \citep{okazaki91-1, papaloizouetal92-1, okazaki97-1}. \cite{hanuschiketal95-1} report line profiles of Fe\,{\textsc{ii}} 5317\,\AA\ (see their Figures 8, 14, and 15) seen in Be stars which are remarkably similar to the Ca\,{\textsc{ii}} triplet observations of SDSS\,J1228+1040.

To test the hypothesis of a fixed intensity structure, we fit the Ca\,{\textsc{ii}} triplet line profile using the method of Doppler tomography. Doppler tomography is commonly employed to deduce the structure of discs in accreting binary systems (\citealt{marsh+horne88-1, steeghs+stehle99-1}, see \citealt{marsh01-2} for a review; in particular his Figure 3). Doppler tomography produces a fixed intensity map (or Doppler map) in velocity space that, for a given period of rotation, best replicates the observed line profiles across all available epochs under the following assumptions \citep{marsh01-2}: 1) All points on the disc are equally visible at all times. 2) The flux from any point fixed in the rotating frame is constant in time. 3) All motion is parallel to the orbital plane. 4) The intrinsic width of the profile from any point is negligible. The Doppler map can also be used to reproduce an emission line profile for a given epoch by reducing the 2D intensity map in velocity space to a 1D intensity distribution in the velocity projected along the corresponding line of sight. Doppler maps are useful tools to study disc structure as the conversion from coordinate space to velocity space is non-unique and multiple intensity profiles across a disc structure can produce the same emission profile. 

We used the 18 Ca\,{\textsc{ii}} triplet profiles collected for SDSS\,J1228+1040 to produce Doppler maps with a range of trial periods spanning 24.64\,years (9000\,days) to 68.45\,years (25000\,days) with increments of 500\,days. We would not expect a period much shorter than $\simeq$\,25 years; otherwise the disc would have precessed $\geq\,180^{\circ}$, producing emission lines that are the reflection of the 2003 March observation.

\begin{figure*}
\centerline{\includegraphics[width=1.8\columnwidth]{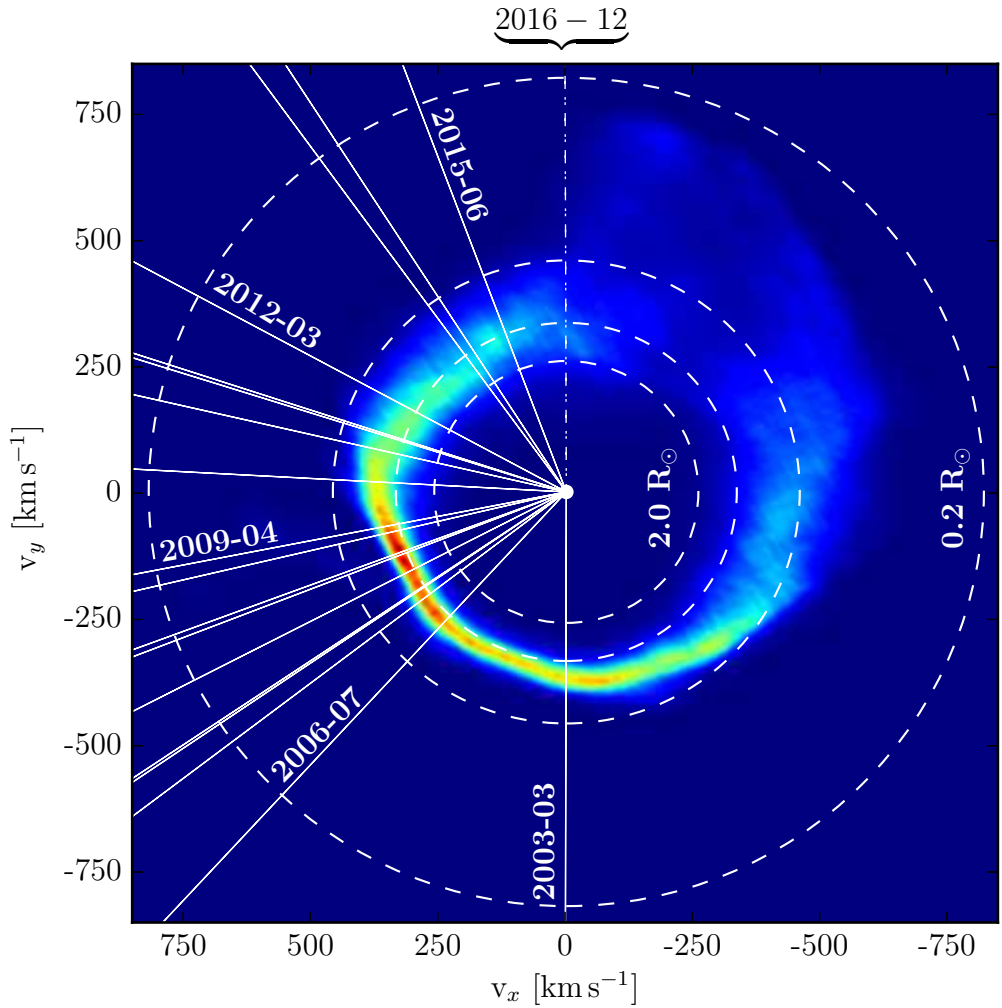}}
\caption{\label{f-dopmap} An intensity distribution in velocity space of the Ca\,{\textsc{ii}} triplet, which models the line profiles observed in SDSS\,J1228+1040 obtained from Doppler tomography. The white dwarf is located at the origin of the map where the lines intersect as a solid white circle. The solid white lines represent the epoch and line of sight for each observation from March 2003 to May 2015 and the dot-dashed white line indicates when we expect the Ca\,{\textsc{ii}} triplet to return to a morphology similar to that observed in 2003 (December 2016). The resulting profile would be a reflection of the first observation, assuming a precession period of $\simeq$\,27\,years. The dashed white circles indicate the location of material in a Keplerian orbit around SDSS\,J1228+1040, with observed orbital radii ($R\sin^{2}i$) of 0.2, 0.64, 1.2 and 2 \Rsun, with the largest circle (highest velocities) corresponding to 0.2\,\Rsun, and the smallest circle (lowest velocities) to 2\Rsun.}
\end{figure*}

\begin{figure}
\centerline{\includegraphics[width=1\columnwidth]{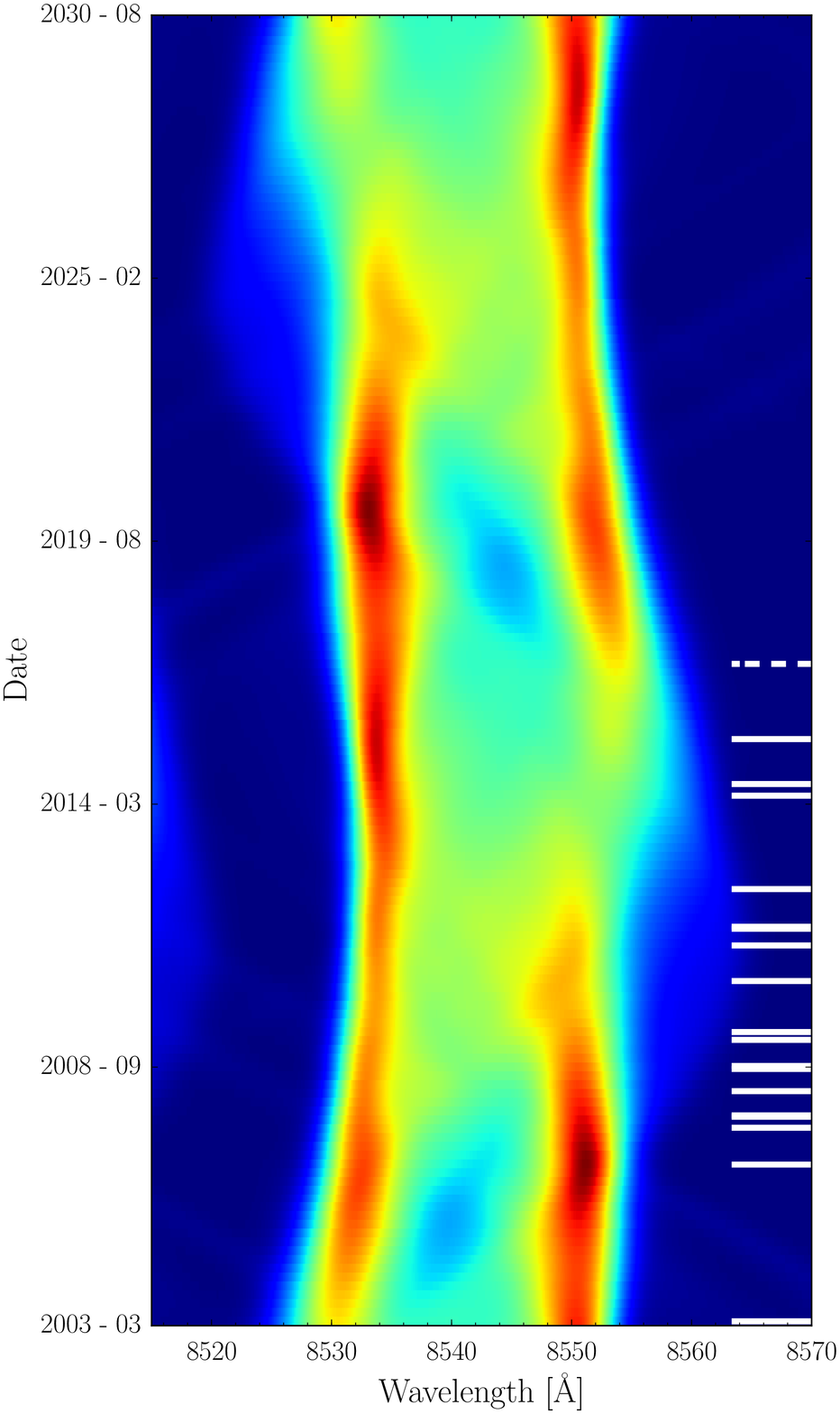}}
\caption{\label{f-doptrail} A trailed spectrogram generated from the Doppler map in Figure\,\ref{f-dopmap} over one full 27.4\,years cycle. The solid and dashed white tabs indicate the position in the spectrogram where we have observed data, and the predicted halfway point respectively.}
\end{figure}

Doppler maps with periods between $\sim$\,24--30\,years ($\sim$\,9000--11000\,days) appear very similar with only slight changes in the shape of the intensity distribution. However, for periods longer than 30\,years the maps of the disc become elongated or stretched, generating discontinuous changes in velocity not expected for Keplerian orbits. Maps with periods longer than 30\,years also generate emission profiles with an additional peak in each of the components of the Ca\,{\textsc{ii}} triplet, which are not observed here or in other systems with Ca\,{\textsc{ii}} emisison (see Section\,\ref{sec:discussion}). We therefore adopt a period in between 24--30\,years; 27.4\,years (10000\,days) and the Doppler map generated with this period is shown in Figure\,\ref{f-dopmap}. We also show a trailed spectrogram generated from the Doppler map over one full 27.4\,years cycle (Figure\,\ref{f-doptrail}). These predictions will be compared against future spectroscopy of the Ca\,{\textsc{ii}} triplet in SDSS\,J1228+1040.

The outer edge of the intensity structure corresponds to the highest velocities and hence the inner edge of the disc in coordinate space. The inner edge is clearly asymmetric, with a large dispersed region and a narrow intense strip related to the large red-shifted velocities and the sharp blue-shifted peak seen in the emission profiles, respectively. 

During the 2003--2008 observations, the dispersed and narrow regions discussed above have projected velocities along the line of sight which are significantly reduced, increasing the central strength near zero velocity rather than the extremes  of the profile. The dashed lines in Figure\,\ref{f-dopmap} represent circular orbits in velocity space, underlining that the Doppler image of SDSS\,J1228+1040 is highly non-circular, which agrees with our qualitative description of the line profiles in Section\,\ref{sec:evolution}. Further, this non-circularity cannot be described well as a single eccentricity, but rather by a changing eccentricity as a function of radius.

From the Doppler map we reconstructed the emission features observed in order to check for self consistency (Figure\,\ref{f-dopmod}). The line profiles generated from the Doppler map fit remarkably well for most epochs, although the sharpness of the red and blue-shifted peaks seen between 2008--2010 are not totally replicated. The inability to fully reconstruct all details of the observed emission profiles may suggest that one or more of the assumptions underlying the method of Doppler tomography are not fulfilled in SDSS\,J1228+1040, e.g. there could be additional short term variability in the system.

An additional note regarding the interpretation of the Doppler map is that it depicts only the intensity distribution of Ca\,{\textsc{ii}}. The difference in morphology between the Ca\,{\textsc{ii}} triplet and other ions, such as O\,{\textsc{i}} (see Section\,\ref{other_emission}) suggests that these elements have different intensity distributions across the disc. Therefore Doppler maps of the emission lines of other ions in combination with that of Ca\,{\textsc{ii}} could reveal additional information on the physical conditions (e.g. density, temperature) within the disc, if time-series spectroscopy with a sufficient signal-to-noise ratio is obtained.

\begin{figure*}
\centerline{\includegraphics[width=2\columnwidth]{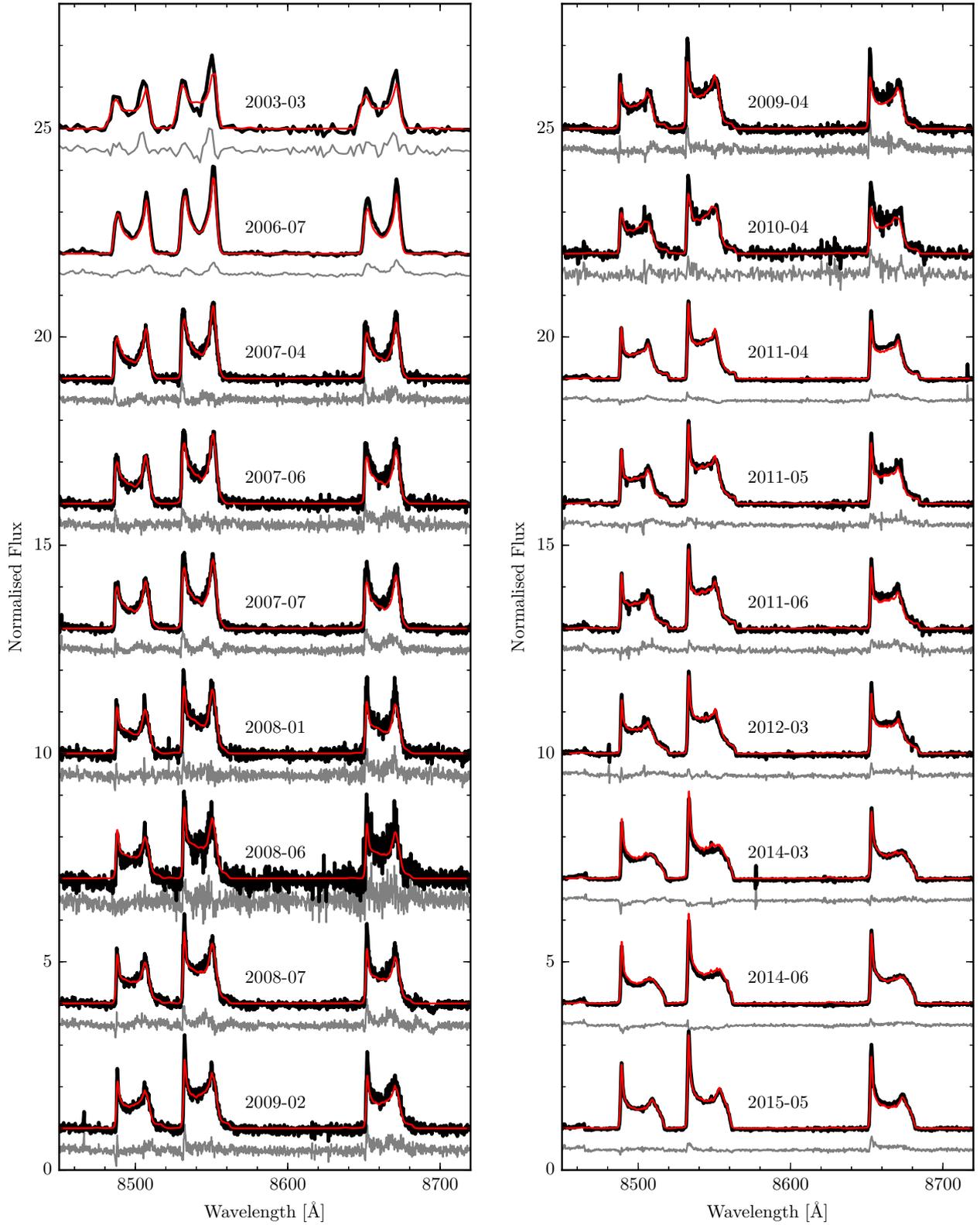}}
\caption{\label{f-dopmod} The continuum-divided Ca\,{\textsc{ii}} triplet line profiles (black) for SDSS\,J1228+1040 in chronological order (as in Figure\,\ref{f-1228_Ca_vel}) reconstructed (red) using the Doppler map in Figure\,\ref{f-dopmap} with residuals (grey). The spectra are shifted in steps of three from the 2009 February (left) and 2015 May observations (right).}
\end{figure*}

\section{A precessing disc?}
\label{sec:disc}

We find that the data and the velocity map shown in Figure\,\ref{f-dopmap} are largely consistent with a constant non-axisymmetric intensity structure that precesses around the white dwarf, resulting in a periodically varying projection into the observed emission line profiles. The precession and the structure of the disc could be explained by several scenarios, which are not necessarily mutually exclusive:

1) We are observing a young, eccentric disc that has only recently formed from the tidal disruption of an asteroid. If only gravitational forces are considered, the asteroid is disrupted and forms a broken, highly-eccentric ring of debris around the white dwarf, the majority of which is located outside of the tidal disruption radius \citep{verasetal14-1}. Adding radiation forces causes the particles in the disc to spread inwards towards the white dwarf (on time-scales determined by their size), eventually circularising \citep{verasetal15-1}. However, the time-scale on which a debris disc circularises is currently poorly known due to the complexity of the forces (e.g. radiation, gas drag) upon and the interactions (e.g. collisions) between the particles. If the circularisation time-scale is significantly shorter than the twelve years spanned by our observations, we would expect the asymmetry in the emission features to have evolved towards a constant, symmetric double-peaked line profile.

2) Precession due to general relativity. General relativistic effects will cause an eccentric orbit to precess over one full orbit on a period $P_\omega$ that is given by

\begin{equation}
P_{\omega} \approx 84.98 \ {\rm yr} 
\left( \frac{M}{0.70{\rm M}_{\odot} } \right)^{-3/2}
\left( \frac{a}{1.00{\rm R}_{\odot}  }  \right)^{5/2}
\left( 1 - e^{2} \right)
\end{equation}

\noindent
where $M$ is the mass of the central star and $a$ is the semi-major axis of the orbit (see \citealt{veras14-1} for details). In the limit of small eccentricities we can ignore the term $(1 - e^{2})$. By adopting \Mwd\,$=0.7$\,\Msun, we find precession periods of 1.54, 27.8 and 134\,years for orbits with semi-major axes of 0.2, 0.64\,\Rsun\ and 1.2\,\Rsun\ respectively, i.e. the period of precession has a strong radial dependence.

General relativistic precession is intrinsic to any orbital mechanics, and furthermore the period of 27.8\,years for an orbit at 0.64\,\Rsun\ is close to the period range deduced from our Doppler maps $\simeq$\,24--30\,years. Therefore it is expected that general relativity should have a significant contribution to the variability observed in this system. The radial dependence of the precession period due to general relativity will cause collisions among the dust particles on different orbits, which is a possible mechanism for generating the observed gaseous component of the debris disc. These collisions would also act to dampen the eccentricity into a circular orbit, where production of gas due to general relativistic precession would eventually cease. It is worth noting that the Doppler map (inherently) does not display the radial dependence of the general relativistic precession, which may be dampened by other forces (e.g. gas pressure effects in the disc).
 
3)  An external perturber outside the disc inducing both the eccentricity and precession. The flyby of a body orbiting on an eccentric orbit would perturb the orbits of the particles in the disc. We have performed some numerical $N$-body simulations to determine the effect of an external perturber on circular rings of particles located at about one Solar radius from the white dwarf.  Because the asymptotic giant branch progenitor of SDSS\,J1228+1040 had a radius of at least 2.5--3.0\,au, surviving planet-sized perturbers must now have a semimajor axis exceeding 3--11\,au (see, e.g. Figure 10 of \citealt{mustill+villaver12-1}).  Consequently, a perturber's current orbital period must be at least several years, which covers multiple observations from Figure\,\ref{f-dopmod}, in particular during the years 2007--2008, where the observations are closely spaced.  We therefore adopted a perturber semimajor axis of 5\,au, with masses ($10^{0} - 10^{4} M_{\oplus}$) and eccentricities (0.990--0.998 - a realistic possibility; see \citealt{debes+sigurdsson02-1, verasetal13-1, voyatzisetal13-1, mustilletal14-1, bonsor+veras15-1, veras+gaensicke15-1}) in ranges that might produce an observable eccentricity change in the ring.  Smaller mass perturbers would have to approach so close to the ring particles to potentially be in danger of tidally disrupting.

Each simulation contained 210 massless ring particles, and spanned one close encounter. The white dwarf mass was set at $0.705\,M_{\odot}$, and the particles were equally partitioned into seven rings at distances of $\left\lbrace 0.6, 0.7, 0.8, 0.9, 1.0, 1.1, 1.2 \,R_{\odot} \right\rbrace$.  We uniformly distributed the initial mean anomalies of the particles within each ring.

In every case changes in ring particle eccentricity occurred only within a few hours of a close encouter. During the remainder of the orbit, there was no eccentricity change.  Therefore, in-between encounters (spanning several years), the geometry of the disc did not change, which is at odds with the observations from Figure\,\ref{f-dopmod}.  This mismatch would also hold true for perturbers with initially greater semimajor axes.
 
4) The disc induces its own eccentricity and precession. \cite{statler01-1} has shown that an eccentric flat fluid disc can precess due to the pressure gradient generated within the disc. Figure 3 from \cite{statler01-1} shows line profiles that are similar in morphology to that of SDSS\,J1228+1040 (see Section\,\ref{sec:evolution}). In contrast to these simulations of purely fluid discs, the gaseous discs that we have observed are coupled to dust grains. \cite{kinnear11} finds a typical gas mass, $M_{\mathrm{gas}}$, of $\sim$\,$10^{19}$\,g. Estimates of the dust mass, $M_{\mathrm{dust}}$, range from $10^{19}$ to $10^{24}$\,g, from disc models that reproduce the observed infrared excess in systems with dusty discs (see, \citealt{juraetal07-1, jura08-1, reachetal09-1}), i.e. $M_{\mathrm{gas}} \leq M_{\mathrm{dust}}$ and as such we can not assume the disc is dominated by the gaseous component. Finally, it is worth noting that \cite{ogilvie01-1} finds that the eccentricity in a disc will dissipate on the viscous time-scale, which we estimate to be $\simeq$\,5-20 years (Section\,\ref{sec:disc}), i.e. similar to our observational baseline. Given that the eccentricity in SDSS\,J1228+1040 persists over this time scale, self-induction seems unlikely.

Of the different scenarios outlined above, a young eccentric disc that precesses due to general relativity appears to be the most natural scenario that can describe our observations. We would also expect this phase to be fairly short-lived, which would explain the rarity of detected gaseous debris disc components. Assuming the adopted precession period of 27.4\,years, the system is expected to reach half way through the precession cycle in December 2016, i.e. we would expect a Ca\,{\textsc{ii}} triplet morphology that is the reflection of the March 2003 observations. In the next section we discuss the variability seen in other gaseous discs around white dwarfs and the possible generation mechanisms for these systems.

\section{Discussion}
\label{sec:discussion}

Besides the observations of SDSS J1228+1040 presented here, changes in the Ca\,{\textsc{ii}} emission lines have been reported for two other systems with a gaseous disc; SDSS\,J0845+2258, and SDSS\,J1617+1620 \citep{wilsonetal14-1, wilsonetal15-1}. In SDSS\,J0845+2257 the Ca\,{\textsc{ii}} profiles evolved in shape from a strong, red-dominated asymmetry to a slight, blue-dominated symmetric profile over a time-scale of ten years; similar to the overall variation seen in the Ca\,{\textsc{ii}} triplet of SDSS\,J1228+1040. In contrast, morphological changes in the line profile were not apparent in SDSS\,J1617+1620, where the Ca\,{\textsc{ii}} emission lines from the gaseous disc remained symmetric (suggesting an axisymmetric intensity pattern) and decreased in strength, eventually disappearing within eight years \citep{wilsonetal14-1}.

There are four other white dwarfs known to host a gaseous disc: SDSS\,J104341.53+085558.2 (henceforth SDSS\,J1043+0855), SDSS\,J073842.56+183509.6 (henceforth SDSS\,J0738+1835), SDSS\,J0959--0200, and HE\,1349--2305. Of these four systems only SDSS\,J1043+0855 and SDSS\,J0738+1835 have multi-epoch spectroscopy. SDSS\,J1043+0855 shows variability similar to SDSS\,J1228+1040 although only a handful of epochs are available so far \citep{manseretal16-1}. In contrast, SDSS\,J0738+1835 shows no change in the shape or strength of the Ca\,{\textsc{ii}} triplet over a period of six years \citep{gaensicke11-1, dufouretal12-1}. However, with only three epochs, more data is needed to confirm the steady state of SDSS\,J0738+1835. The emission lines in HE\,1349--2305 show a large asymmetry similar to the systems with a morphological variation (SDSS\,J1228+1040, SDSS\,J1043+0855 and SDSS\,J0845+2257), and therefore this star would be an ideal candidate for follow up observations to detect variability \citep{melisetal12-1}. The Ca\,{\textsc{ii}} triplet observed at SDSS\,J0959--0200 is much narrower and weaker than that of the other six systems, indicating a low inclination of the disc with respect to the line of sight \citep{farihietal12-1}. Finally, \cite{guoetal15-1} recently identified the candidate gas disc at SDSS\,J114404.74+052951.6 which has Ca\,{\textsc{ii}} emission similar that of SDSS\,J0959--0200. The system also shows infrared excess attributed to a dusty debris disc, but the Ca\,{\textsc{ii}} emission lines are extremely weak, and follow up observations are needed to confirm the presence of a gaseous component.

The gaseous component of a debris disc will accrete onto the white dwarf, and without a sustained generation mechanism will dissipate over the viscous time-scale, which can be estimated using equation 2 from \cite{metzgeretal12-1}. The major uncertainty is the value of $\alpha$, the accretion disc viscosity parameter \citep{shakura+sunyaev73-1}. \cite{kingetal07-1} suggest $0.1 \leqslant \alpha \leqslant 0.4$ for a thin fully ionised disc. Given that the metallic gas is ionised to a high degree, it is plausible to adopt this range for $\alpha$, which leads to viscous time-scales of 5--20\,years for a typical white dwarf mass. These time-scales match the disappearance of the Ca\,{\textsc{ii}} lines observed in SDSS\,J1617+1620 discussed above.

\begin{table}
\centering
\caption{Equivalent width measurements of the Ca\,{\textsc{ii}} triplet in SDSS\,J1228+1040. \label{t-ewtrip}}
\begin{tabular}{lr}
\hline
Date & Equivalent width [\AA] \\
\hline
2003--03 & -60 (7)\\
2006--07 & -69 (2)\\
2007--04 & -61 (2)\\
2007--06 & -63 (2)\\
2007--07 & -63 (2)\\
2008--01 & -60 (2)\\
2008--06 & -54 (2)\\
2008--07 & -60 (2)\\
2009--02 & -58 (2)\\
2009--04 & -62 (2)\\
2010--04 & -67 (1)\\
2011--01 & -57 (1)\\
2011--05 & -60 (1)\\
2011--06 & -59 (1)\\
2012--03 & -56 (1)\\
2014--03 & -50 (1)\\
2014--06 & -52 (1)\\
2015--05 & -57 (1)\\
\hline
\end{tabular}
\end{table}

We calculated the equivalent width of the Ca\,{\textsc{ii}} triplet for SDSS\,J1228+1040 for each epoch (Table\,\ref{t-ewtrip}), and do not detect any significant decrease, suggesting that; 1) the gaseous discs in SDSS\,J1228+1040 and SDSS\,J0845+2257 are being sustained by ongoing gas production, and 2) there may be multiple pathways to generate a gaseous disc given that there are clear differences in the type of variability seen in these three systems and SDSS\,J1617+1620. 

From Section\,\ref{sec:disc}, a natural scenario to explain the morphology variation is that of a young eccentric disc that precesses due to general relativity with the radial dependence of the precession period resulting in ongoing collisional gas production that ceases once the disc circularises. This scenario allows for gas generation, while not being mutually exclusive with the observations of SDSS\,J1617+1620, which would be better explained by a transient event such as a secondary asteroid impact onto a pre-existing, circularised disc \citep{jura08-1}. Such an asteroid impact would generate an initial quantity of gas that would subsequently dissipate over the viscous time-scale. 

The relatively short-lived transient nature of these gas generation mechanisms also explains why there are systems with accretion rates much larger than that of SDSS\,J1228+1040, such as PG\,0843+516 and GALEX\,1931+0117, which do not show any sign of a gaseous disc \citep{gaensickeetal12-1}. We expect that debris discs with no detectable gaseous component are largely circularised.

\section{Conclusions}
\label{sec:conc}

We report the pronounced morphology changes of the Ca\,{\textsc{ii}} triplet at SDSS\,J1228+1040 which we have modelled using the method of Doppler tomography. This procedure produced a non-axisymmetric intensity map in velocity space of the Ca\,{\textsc{ii}} triplet, which we have interpreted as a precessing disc with a period in the range 24--30\,years. We also detected additional emission lines in the averaged X-shooter spectra of SDSS\,J1228+1040, increasing the number of observed gaseous elements in this system to four. The variation in shape between the emission features of different ions is a clear indication that the intensity distribution in the disc for each ion is not the same. With time-resolved spectroscopy and sufficient signal-to-noise, Doppler tomography of the additional line profiles (such as O\,{\textsc{i}} 7774, 8446\,\AA) may reveal information about the physical properties in the disc (e.g density and temperature distributions). 

The gaseous component of debris discs are a tracer of dynamical activity in these systems. Four of the five gaseous discs with time-resolved observations over time-scales of years show variability, and small-number statistics so far suggest that the dynamic variability of the gaseous discs themselves is the the norm rather than the exception. 

\section*{Acknowledgements}
We thank the referee for their helpful comments and suggestions for improving this manuscript. The research leading to these results has received funding from the European Research Council under the European Union's Seventh Framework Programme (FP/2007-2013) / ERC Grant Agreement n.\,320964 (WDTracer). We would like to thank Yan-Ping Chen and Scott Trager for sharing their X-shooter telluric template library. TRM acknowledges support from the Science and Technology Facilities Council (STFC), grant ST/L000733/1. SGP acknowledges financial support from FONDECYT in the form of grant number 3140585. CJM would like to thank Mark Hollands for his help with Figure\,\ref{f-1228_plot}. 
 
Based on observations made with ESO Telescopes at the La Silla Paranal Observatory under programme IDs: 079.C-0085, 081.C-0466, 082.C-0495, 087.D-0139, 088.D-0042, 093.D-0426, 095.D-0802, 192.D-0270, 383.C-0695, 386.C-0218, 595.C-0650. This work has made use of observations from the SDSS-III, funding for which has been provided by the Alfred P. Sloan Foundation, the Participating Institutions, the National Science Foundation, and the U.S. Department of Energy Office of Science. The SDSS-III web site is http://www.sdss3.org/. Based on observations made with the NASA/ESA HST, obtained at the Space Telescope Science Institute, which is operated by the Association of Universities for Research in Astronomy, Inc., under NASA contract NAS 5-26555. These observations are associated with program 11561. Based on observations made with the WHT operated on the island of La Palma by the Isaac Newton Group in the Spanish Observatorio del Roque de los Muchachos of the Instituto de Astrofísica de Canarias.

\bibliographystyle{mnras}
\bibliography{aamnem99,aabib}

\bsp

\label{lastpage}

\end{document}